\documentclass[notitlepage,letterpaper,showpacs,preprintnumbers,amsmath,nofootinbib,amssymb,twocolumn, superscriptaddress, hyperref]{revtex4}
\pdfoutput=1
\usepackage{amssymb,amsmath,latexsym,mathrsfs}
\usepackage{graphicx}
\usepackage{psfrag}
\usepackage{braket}
\usepackage[usenames,dvipsnames]{color}
\usepackage[colorlinks=true, linkcolor=BrickRed, citecolor=Blue, urlcolor=Blue, filecolor=Blue, draft]{hyperref}
 \usepackage[normalem]{ulem}
\usepackage{amsmath}
\usepackage{slashed}
\usepackage{mathtools}
\usepackage[titletoc]{appendix}
\usepackage{datetime}
\usepackage{alltt} 
\usepackage{esvect}
\usepackage{epstopdf}

\newcommand{\beq}{\begin{equation}}
\newcommand{\eeq}{\end{equation}}
\newcommand{\ben}{\begin{eqnarray}}
\newcommand{\een}{\end{eqnarray}}
\newcommand{\bi}{\begin{itemize}}
\newcommand{\ei}{\end{itemize}}

\begin{document}

\title{Can interacting dark energy solve the $H_0$ tension?}  

\author{Eleonora Di Valentino}
\email{eleonora.di\_valentino@iap.fr}
\affiliation{Institut d'Astrophysique de Paris (UMR7095: CNRS \& UPMC- Sorbonne Universities), F-75014, Paris, France}
\affiliation{Sorbonne Universit\'es, Institut Lagrange de Paris (ILP), F-75014, Paris, France}
\author{Alessandro Melchiorri}
\email{alessandro.melchiorri@roma1.infn.it}
\affiliation{Physics Department and INFN, Universit\`a di Roma ``La Sapienza'', Ple Aldo Moro 2, 00185, Rome, Italy}
\author{Olga Mena}
\email{omena@ific.uv.es}
\affiliation{IFIC, Universidad de Valencia-CSIC, 46071, Valencia, Spain}

\preprint{}
\begin{abstract}
The answer is Yes! We indeed find that interacting dark energy can alleviate the current tension
on the value of the Hubble constant $H_0$ between 
the Cosmic Microwave Background anisotropies constraints obtained from the Planck satellite and the recent direct
measurements reported by Riess et al. 2016. The combination of these two datasets points towards an evidence for a non-zero dark matter-dark energy coupling $\xi$ at
more than two standard deviations, with $\xi=-0.26_{-0.12}^{+0.16}$ at $95\%$ CL.
However the $H_0$ tension is better solved when the equation of
state of the interacting dark energy component is allowed to freely vary, with a phantom-like equation of state $w=-1.184\pm0.064$ (at $68
\%$ CL), ruling out the pure cosmological constant case, $w=-1$, again at more than two standard deviations. 
When Planck data are combined with external datasets, as BAO, JLA
Supernovae Ia luminosity distances, cosmic shear or lensing data, we find perfect consistency with the cosmological constant scenario and no compelling 
evidence for a dark matter-dark energy coupling. 
\end{abstract}
\maketitle

\section{Introduction}
The recent measurements of Cosmic Microwave Background anisotropies
(CMB) provided by the Planck satellite have fully confirmed the predictions of the standard cosmological model (hereafter, $\Lambda$CDM), based on cold dark matter, a cosmological constant and inflation~\cite{planckparams2015}. However, when constraints on the cosmological
parameters of the $\Lambda$CDM model are derived from the Planck data, some tensions appear between
their values and the corresponding values obtained from independent, complementary, observables.

The most important tension concerns the value of the Hubble constant. Indeed, the latest analysis of 
CMB temperature and polarization data from the Planck experiment provides the constraint 
of $H_0=66.93 \pm 0.62$ km/s/Mpc at $68 \%$ CL, obtained assuming $\Lambda$CDM ~\cite{newtau}.
This is more than $3 \sigma$ away from the recent direct and local determination of Riess et al. 2016 (R16,
hereafter) of $H_0=73.24\pm1.74$ km/s/Mpc~\cite{R16}.

Another important discrepancy is present on the recovered values of the $S_8=\sigma_8 \sqrt{\Omega_m/0.3}$ parameter (where
$\sigma_8$ is the amplitude of matter fluctuations and $\Omega_m$ is the matter density)
derived independently by Planck  and weak lensing surveys such as CFHTLenS~\cite{Heymans:2012gg,Erben:2012zw} and KiDS-450~\cite{Hildebrandt:2016iqg}. Considering 
a $\Lambda$CDM scenario,  the KiDS-450 result is in tension with the Planck constraint at about $2.3$ standard deviations (see
e.g.~\cite{Joudaki:2016kym}). 

Clearly, unresolved systematics can still play a key role in explaining these discrepancies, however
several physical mechanisms beyond $\Lambda$CDM that could change the
derived values of $H_0$ and/or $S_8$ from Planck have been proposed, either solving or alleviating the tensions with the
extracted values from R16 and cosmic shear local measurements (see
e.g. Refs.~\cite{dms0,dms,bernal,dmnu1,dmnu2,decay,Gariazzo:2017pzb,
Grandis:2016fwl,Zhao:2017urm,Yang:2017amu,Prilepina:2016rlq,DiValentino:2016ziq,Santos:2016sog,Kumar:2016zpg,Joudaki:2016kym,Karwal:2016vyq,Ko:2016uft,Archidiacono:2016kkh,Qing-Guo:2016ykt,Chacko:2016kgg,Zhang:2017idq,zhao,Sola:2017jbl,brust,Lancaster:2017ksf}).

In this paper we focus our attention on dark energy: it has indeed been shown that
introducing a dark energy equation of state (constant with time) $w<-1$ not only can
solve the tension on the Hubble parameter but also does it in a more
efficient way than other non-standard extensions as, for example, the inclusion of extra relativistic degrees of freedom, 
via the $N_{eff}$ parameter~\cite{dms}.
At the same time, a dynamical dark energy component seems to be favoured in combined
analyses of Planck and cosmic shear data (see e.g. \cite{Joudaki:2016kym}).

On top of that we should not forget that a cosmological constant clearly presents a puzzling and controversial 
solution to the dark energy problem, being probably the major theoretical 
weakness of the standard cosmological model. The possibility of having a different candidate for 
this component should be therefore investigated and any hints
for deviations from the $\Lambda$CDM picture must be carefully scrutinized.

If the solution for the current tensions is within the dark energy sector then it is worthwhile to 
investigate which dark energy model is better suited for this task.
Recently, several authors have considered different parameterizations
of the dark energy equation of state, deriving bounds arising from a
number of cosmological datasets (see e.g. \cite{Zhao:2017urm,dms02}).
Here we focus on interacting dark matter-dark energy models. Indeed,
 a larger value for the Hubble parameter from CMB 
data can be obtained by including possible interactions between the dark energy
and dark matter components  (see e.g. \cite{Salvatelli,Sola1,Sola2,Saulo,Richarte,Valiviita,Elisa,Murgia:2016ccp}).
More specifically, assuming the interacting dark energy 
model presented in Ref.~\cite{ideolga}, when considering the Planck 2013
data release, one gets the constraint $H_0=72.1_{-2.3}^{+3.2}$
km/s/Mpc at $68 \%$ CL, that is significantly larger than the constraint $H_0=67.3\pm1.2$ km/s/Mpc at $68 \%$ CL 
obtained with the same dataset but assuming a cosmological constant and no interaction (see Ref.~\cite{ide2}).

 It is therefore timely to investigate if the same interacting dark energy model 
 can also solve the tension between the new Planck 2015 data release (that includes
 new polarization data, which significantly improve the constraining
 power of these measurements) and the new bound R16, that in practice reduces by $\sim 42 \%$ the 
 uncertainty of previous late-universe constraints on the Hubble constant~\cite{riess11}.

In what follows we perform such an analysis, showing that indeed the
current tension on $H_0$ can be solved by introducing an interaction between
dark energy and dark matter as the one proposed by Ref.~\cite{ideolga}.
 Moreover, it is also important to examine if the coupling can solve the 
tension {\it more efficiently} than a dark energy equation of state $w<-1$.
For this reason we also perform an analysis by varying at the same time both
the coupling and the equation of state of the dark energy component,
as in interacting scenarios the dark energy equation of state of the
dark matter fluid could in principle be different
from that corresponding to the cosmological constant case.

Our work is organized as follows: in the following section we describe the interacting
dark energy model assumed here, in Sec.~III we describe the cosmological
data and the analysis method and in Sec.~IV we show the obtained results.
Finally, we draw our conclusions in Sec.~V.

\section{Interacting Dark Energy}

As previously stated, we consider the interacting dark energy scenario
of Refs.~\cite{ideolga,ide2}, which can be parameterized as:
\begin{eqnarray}
  \label{eq:DM}
\nabla_\mu T^\mu_{(dm)\nu} &=&Q \,u_{\nu}^{(dm)}/a~, \\
  \label{eq:DE}
\nabla_\mu T^\mu_{(de)\nu} &=&-Q \,u_{\nu}^{(dm)}/a~.
\end{eqnarray}
In the equations above, $T^\mu_{(dm)\nu}$ ($T^\mu_{(de)\nu}$) refers to
the dark matter (dark energy) energy-momentum tensor, $Q$ is
the interaction rate and $u_{\nu}^{(dm)}$ is the four-velocity of the
dark matter fluid. In order to avoid instabilities in the evolution of
the linear perturbations, we restrict ourselves to the case in which $Q$ reads
as:
\begin{equation}
Q=\xi \mathcal{H} \rho_{de}~,
\label{eq:rate}
\end{equation}
i.e. the interaction rate is proportional to the dark energy density
$\rho_{de}$ via a dimensionless parameter $\xi$ (that needs to be
negative) and $\mathcal{H}=\dot{a}/a$, with the dot referring to derivative respect
to conformal time.

The evolution equations for the interacting background read as~\cite{ideolga}
\begin{eqnarray}
  \label{eq:bDM}
\dot{{\rho}}_{dm}+3{\mathcal H}{\rho}_{dm}= \xi{\mathcal H}{\rho}_{de}~, \\
  \label{eq:bDE}
\dot{{\rho}}_{de}+3{\mathcal H}(1+w){\rho}_{de}= -\xi{\mathcal
  H}{\rho}_{de}~.
\end{eqnarray}  

The perturbation evolution, within the linear regime, and in the
synchronous gauge, is given by~\cite{ideolga}
 \begin{eqnarray}
\label{eq:delbe}
\dot\delta_{dm}  & = & -(k v_{dm}+\frac12 \dot h) +\xi {\mathcal H}\frac{\rho_{de}}{\rho_{dm}} \left(\delta_{de}-\delta_{dm}\right)
\\ \nonumber
&& +\xi \frac{\rho_{de}}{\rho_{dm}} \left(\frac{k v_T}{3}+\frac{\dot h}{6}\right)\,, \\
\label{eq:deles}
\dot\delta_{de}  & = & -(1+w)(k v_{de}+\frac12 \dot h)-3 {\mathcal H}\left(1 -w\right)
   \\ \nonumber
&& \left[ \delta_{de} +{\mathcal H} \left( 3(1+w) + \xi\right)\frac{v_{de}}{k} \right]-\xi \left(\frac{k v_T}{3}+\frac{\dot h}{6}\right) \,,\\
\label{eq:thes}
 \dot v_{dm}  & = & -{\mathcal H} v_{dm} \,,\\
 \label{eq:thees}
\dot v_{de}  & = & 2 {\mathcal H}\left(1 +\frac{\xi}{1+w} \right)
    v_{de}+\frac{k}{1+w}\delta_{de}-\xi{\cal H}\frac{v_{dm}}{1+w}\,,
\end{eqnarray}
with $\delta_{dm,de}$ and $v_{dm,de}$ the density perturbation
and the velocity of the two fluids, $v_T$ is the center of mass velocity for the total
fluid and the dark energy speed of sound is $\hat
c_{s\,de}^2=1$. The equations above include the contributions of the perturbation in the expansion rate
$H={\mathcal H}/a + \delta H$. 

Following Refs.~\cite{ideolga,Doran:2003xq,Majerotto:2009np}, we have considered adiabatic initial
conditions for all components. It has been shown there that, if one
assumes adiabatic initial conditions for all the standard cosmological fluids (photon,
baryons,...), the coupled dark energy fluid will also obey adiabatic
initial conditions. In the synchronous gauge
and at leading order in $x=k\tau$, the initial conditions read:
\begin{widetext}
\begin{eqnarray}
  \delta_{de}^{in} (x)&=&(1+w-2\xi)\frac{(1+w+\xi/3)}{12w^2-2w-3w \xi+7
    \xi-14}\, 
  \left(\frac{-2\delta_\gamma^{in}(x)}{1+w_\gamma}\right)\,, \cr
  v_{de}^{in}&=&\frac{x(1+w+\xi/3)}{12w^2-2w-3w \xi+7
    \xi-14}\,
  \left(\frac{-2\delta_\gamma^{in}(x)}{1+w_\gamma}\right)\,, \nonumber
\end{eqnarray}
\end{widetext}
where $\delta_\gamma^{in}(x)$ are the initial conditions for the
photon density perturbations and $w_\gamma=1/3$ is the equation of
state of the photon, see Ref.~\cite{Ballesteros:2010ks} for the uncoupled case.

\section{Method}

The interacting dark energy scenario requires the six standard
cosmological parameters of the $\Lambda$CDM plus one more parameter
defined in the previous section, the coupling $\xi$. In particular,
for the $\Lambda$CDM model, the parameters are the baryon density
$\Omega_bh^2$, the cold dark matter density $\Omega_ch^2$, the
reionization optical depth $\tau$ and the ratio between the sound
horizon and the angular diameter distance at decoupling $\theta_{s}$. 
Furthermore, we consider two parameters directly related to the
inflationary paradigm, that are the
spectral index $n_S$ and the logarithm of the amplitude of the primordial power spectrum, $ln(10^{10}A_S)$.
As a second step, we extend this baseline model, by adding one more
parameter, a freely varying dark energy equation of state $w$, assumed
to be constant in redshift.

We analyze this scenario by combining several cosmological probes.
We consider the full temperature power spectrum provided by the Planck 
collaboration~\cite{Aghanim:2015xee} at multipoles
$2\leq\ell\leq2500$, in combination with the low-$\ell$ polarization power spectra in the multipoles range $2\leq\ell\leq29$.
We refer to this combination as ``Planck TT + lowTEB''.
We also include the high multipole Planck polarization
spectra~\cite{Aghanim:2015xee}, in the multipole range
$30\leq\ell\leq2500$. We refer to this combination as ``Planck TTTEEE + lowTEB''. However, we would like to remind here that this
combination of datasets is considered less robust as it still under discussion due to some possible residual
systematics contamination~\cite{Aghanim:2015xee,planckparams2015}.

Additionally to the CMB datasets described above, we consider their combination with the following cosmological measurements:
\begin{itemize}
\item \textbf{tau055}: We replace the ``lowTEB'' Planck data with a gaussian prior on the reionization optical depth $\tau=0.055\pm0.009$, as obtained from the Planck HFI measurements in~\cite{newtau};
\item \textbf{lensing}: We consider the 2015 Planck CMB lensing
  reconstruction power spectrum $C^{\phi\phi}_\ell$ obtained with the
  CMB trispectrum analysis~\cite{Ade:2015zua};
 \item \textbf{BAO}: Baryon Acoustic Oscillation measurements from the
    6dFGS~\cite{beutler2011}, SDSS-MGS~\cite{ross2014},
    BOSSLOWZ and CMASS-DR11~\cite{anderson2014}
    surveys are also considered;
\item \textbf{R16}: As previously stated, we include a gaussian prior on the Hubble constant
  $H_0=73.24\pm1.74$ km/s/Mpc,
 by quoting the value provided with direct measurements of luminosity distances in Riess et al.~\cite{R16};
\item \textbf{JLA}: We also employ luminosity distance data of Supernovae type Ia from the "Joint Light-curve Analysis"
derived from the SNLS and SDSS catalogs~\cite{JLA};
\item \textbf{WL}: We add weak lensing galaxy data from the
  CFHTlens~\cite{Heymans:2012gg,Erben:2012zw} survey with the priors 
and conservative cuts to the data as described in Ref.~\cite{planckparams2015}.
\end{itemize}

The analysis is done with a modified version of the most recent publicly available Monte-Carlo Markov Chain package \texttt{cosmomc}~\cite{Lewis:2002ah}, with a convergence diagnostic based on the Gelman and Rubin statistics. As the original code, this version implements an efficient sampling of the posterior distribution using the fast/slow parameter decorrelations~\cite{Lewis:2013hha}, and it includes the support for the Planck data release 2015 Likelihood Code~\cite{Aghanim:2015xee} (see \url{http://cosmologist.info/cosmomc/}).

\section{Results}

\begin{table*}
\begin{center}\footnotesize
\scalebox{0.8}{\begin{tabular}{c|ccccccc}
Parameter         &Planck TT& Planck TT& Planck TT & Planck TT & Planck TT & Planck TT & Planck TT  \\            
 & + lowTEB & + lowTEB + R16 & + lowTEB + BAO& + lowTEB + JLA & + lowTEB + WL & + lowTEB + lensing & + tau055 \\                 
\hline
\hspace{1mm}\\
$\Omega_bh^2$  &  $0.02222\,\pm0.00023$& $0.02235\pm0.00022$  & $0.02221\pm0.00020$& $0.02220\pm0.00023$ &  $0.02232\,\pm0.00023$& $0.02225\pm0.00023$  & $0.02207\pm0.00021$ \\
\hspace{1mm}\\
$\Omega_ch^2$  &  $0.100\,^{+0.017}_{-0.011}$&  $0.088\,^{+0.006}_{-0.013}$&  $ 0.104\,^{+0.015}_{-0.007}$&  $ 0.104\,^{+0.014}_{-0.007}$ &  $0.098\,^{+0.016}_{-0.011}$&  $0.099\,^{+0.017}_{-0.011}$&  $0.098\,^{+0.012}_{-0.018}$\\
\hspace{1mm}\\
$\tau$ &   $0.076\,\pm0.019$& $0.082\,\pm0.019$& $0.076\pm0.0018$ &   $0.076\,\pm0.019$& $0.075\,\pm0.019$&  $0.065\,\pm0.016$& $0.0587\pm0.0089$ \\
\hspace{1mm}\\
$n_s$ &  $0.9652\,\pm0.0061$  & $0.9696\,\pm0.0059$&$0.9650\,\pm0.0047$ &  $0.9648\,\pm0.0059$& $0.9691\,\pm0.0060$  & $0.9675\,\pm0.0059$&$0.9585\,\pm0.0058$  \\
\hspace{1mm}\\
$ln(10^{10}A_s)$ &  $3.087\,\pm0.036$& $3.095\,\pm0.037$& $3.087\,\pm0.035$ & $3.087\,\pm0.036$& $3.080\,\pm0.037$& $3.061\,\pm0.029$& $3.057\,\pm0.018$\\
\hspace{1mm}\\
$H_0 [\rm{km \, s^{-1} \, Mpc^{-1}}]$ &$69.0\,\pm1.4$&$70.7\,^{+1.2}_{-1.0}$ &$68.63\,^{+0.8}_{-1.0}$ &$68.6\,\pm1.2$ &$69.8\,\pm1.4$&$69.5\,\pm1.4$ &$68.4\,^{+1.6}_{-1.4}$  \\
\hspace{1mm}\\
$\sigma_8$ &$1.01\,_{-0.19}^{+0.08}$& $1.13\,_{-0.10}^{+0.18}$& $0.967\,^{+0.05}_{-0.15}$ &$0.964\,_{-0.14}^{+0.05}$& $1.01\,^{+0.09}_{-0.19}$  & $1.00\,_{-0.19}^{+0.09}$& $1.04\,^{+0.12}_{-0.22}$ \\
\hspace{1mm}\\
$\Omega_m$ & $0.260\,\pm0.036$  & $0.223\,^{+0.015}_{-0.032}$& $0.270\,^{+0.038}_{-0.021}$ &$0.272\,_{-0.025}^{+0.036}$& $0.249\,\pm0.033$  & $0.252\,\pm0.035$& $0.260\,^{+0.033}_{-0.051}$ \\
\hspace{1mm}\\
$\xi$  &$>-0.232$&     $-0.25\,^{+0.05}_{-0.10}$&$>-0.184$  &   $>-0.181$&$-0.17\,^{+0.12}_{-0.10}$&            $>-0.232$&$-0.21\,^{+0.09}_{-0.16}$\\
\hline
\end{tabular}}
\end{center}
\caption{$68\%$~CL constraints on cosmological parameters from the 
  ``Planck TT + lowTEB'' baseline dataset and its further combinations with several datasets (see text). An interacting
  dark energy model with negative coupling $\xi\le0$ is assumed. Please note that in case of the "tau055" prior the "lowTEB" dataset is not considered.}
\label{tab:temp}
\end{table*}

\begin{table*}
\begin{center}\footnotesize
\scalebox{0.8}{\begin{tabular}{c|ccccccc}
Parameter         & Planck TTTEEE & Planck TTTEEE & Planck TTTEEE & Planck TTTEEE & Planck TTTEEE &  Planck TTTEEE & Planck TTTEEE   \\            
 & + lowTEB & + lowTEB + R16 & + lowTEB + BAO& + lowTEB + JLA & + lowTEB + WL & + lowTEB + lensing & + tau055 \\                 
\hline
\hspace{1mm}\\
$\Omega_bh^2$  &  $0.02223\,\pm0.00016$& $0.02230\pm0.00015$  & $0.02223\pm0.00014$& $0.02222\pm0.00015$ &  $0.02228\,\pm0.00016$& $0.02224\pm0.00016$  & $0.02214\pm0.00015$ \\
\hspace{1mm}\\
$\Omega_ch^2$  &  $0.101\,^{+0.017}_{-0.010}$&  $0.089\,^{+0.005}_{-0.012}$&  $ 0.104\,^{+0.014}_{-0.007}$&  $ 0.104\,^{+0.014}_{-0.007}$ &  $0.099\,^{+0.017}_{-0.011}$&  $0.097\,^{+0.018}_{-0.010}$&  $0.097\,^{+0.011}_{-0.018}$\\
\hspace{1mm}\\
$\tau$ &   $0.077\,\pm0.017$& $0.081\,\pm0.017$& $0.078\pm0.0016$ &   $0.078\,\pm0.017$& $0.074\,\pm0.017$&  $0.062\,\pm0.014$& $0.0602\pm0.0087$ \\
\hspace{1mm}\\
$n_s$ &  $0.9641\,\pm0.0047$  & $0.9664\,\pm0.0048$&$0.9643\,\pm0.0043$ &  $0.9641\,\pm0.0047$& $0.9660\,\pm0.0048$  & $0.9652\,\pm0.0047$&$0.9593\,\pm0.0046$  \\
\hspace{1mm}\\
$ln(10^{10}A_s)$ &  $3.090\,\pm0.033$& $3.095\,\pm0.033$& $3.091\,\pm0.032$ & $3.091\,\pm0.033$& $3.082\,\pm0.033$& $3.056\,\pm0.025$& $3.059\,\pm0.018$\\
\hspace{1mm}\\
$H_0 [\rm{km \, s^{-1} \, Mpc^{-1}}]$ &$68.9\,\pm1.2$&$70.2\,^{+1.1}_{-0.8}$ &$68.58\,^{+0.8}_{-1.1}$ &$68.57\,^{+0.9}_{-1.2}$ &$69.3\,\pm1.2$&$69.2\,\pm1.2$ &$68.7\,^{+1.4}_{-1.3}$  \\
\hspace{1mm}\\
$\sigma_8$ &$1.01\,_{-0.19}^{+0.08}$& $1.14\,_{-0.09}^{+0.017}$& $0.966\,^{+0.05}_{-0.15}$ &$0.967\,_{-0.14}^{+0.05}$& $1.01\,^{+0.09}_{-0.19}$  & $0.99\,_{-0.19}^{+0.08}$& $1.05\,^{+0.12}_{-0.21}$ \\
\hspace{1mm}\\
$\Omega_m$ & $0.261\,^{+0.044}_{-0.035}$  & $0.227\,^{+0.014}_{-0.031}$& $0.271\,^{+0.039}_{-0.021}$ &$0.271\,_{-0.023}^{+0.038}$& $0.254\,_{-0.036}^{+0.041}$  & $0.257\,_{-0.034}^{+0.044}$& $0.255\,^{+0.030}_{-0.049}$ \\
\hspace{1mm}\\
$\xi$  &$>-0.229$&     $-0.259\,^{+0.043}_{-0.098}$&$>-0.182$  &   $>-0.182$&$-0.18\,^{+0.13}_{-0.10}$&            $>-0.230$&$-0.21\,^{+0.08}_{-0.15}$\\
\hline
\end{tabular}}
\end{center}
\caption{$68\%$~CL constraints on cosmological parameters from the 
  ``Planck TTTEEE + lowTEB'' baseline dataset and its further combinations with several datasets (see text). An interacting
  dark energy model with negative coupling $\xi\le0$ is assumed. Please note that in case of the "tau055" prior the "lowTEB" dataset is not considered.}\label{tab:pol}
\end{table*}

\begin{figure*}[t]
\begin{center}
\includegraphics[width=.4\textwidth]{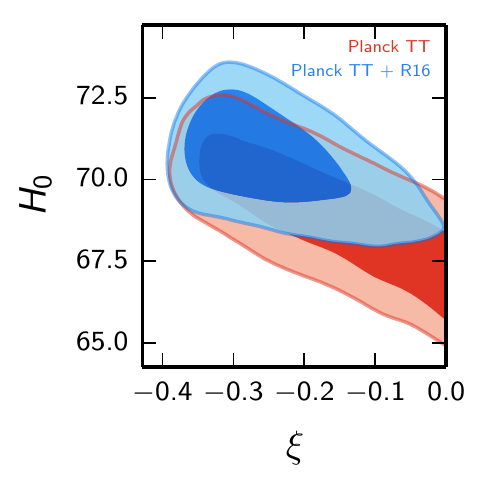}\includegraphics[width=.4\textwidth]{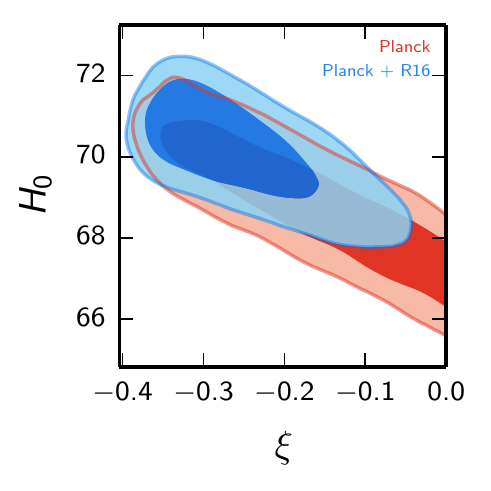} 
\caption{$68\%$ and $95\%$ CL in the two-dimensional 
($\xi$, $H_0$) planes from the ``Planck TT + lowTEB'' dataset (left panel) and ``Planck TTTEEE +
lowTEB'' dataset (right panel) also combined with the R16 prior on the
Hubble constant. Notice that the presence of a coupling $\xi$ allows
for larger values for $H_0$ from Planck data. The inclusion of the R16 prior results
in an indication for $\xi<0$ with a significance above two standard deviations.}
\label{fig1}
\end{center}
\end{figure*}

The constraints at $68 \%$~CL on the cosmological parameters obtained 
by including the interaction between dark matter and
dark energy  are given in Tabs.~\ref{tab:temp} and \ref{tab:pol} for
the ``Planck TT + lowTEB'' and the ``Planck TTTEEE + lowTEB'' baseline
datasets, respectively. Those obtained allowing also for a freely-varying dark
energy equation of state are shown in Tabs.~\ref{tab:temp_w} and
\ref{tab:pol_w}. For comparison purposes we also quote in
Tab.~\ref{tab:pol_noint} the bounds on the cosmological parameters
 as obtained from the Planck collaboration~\cite{planckparams2015} in
 the pure $\Lambda$CDM and $w$CDM context, i.e. without considering the dark
 matter-dark energy coupling $\xi$.

Notice, first of all, that the CMB-only constraints on the coupling $\xi$ are
very similar with or without the inclusion of the polarization data
from Planck at higher multipoles. Tables~\ref{tab:temp} and
\ref{tab:pol} show that, in the $\Lambda$CDM + $\xi$ scenario, CMB
data only imposes a lower limit in the coupling $\xi>-0.23$ at $68 \%$
CL, with or without the Planck small-scale polarization data.

However, if we compare the CMB constraints on the Hubble constant in Tabs.~\ref{tab:temp} and
\ref{tab:pol} in the presence of a coupling $\xi$ to those obtained with no interaction in Tab.~\ref{tab:pol_noint},
we see that the coupling produces a shift at more than $2\sigma$ towards higher values of the Hubble 
constant and relaxes the error bars by a factor $\sim 2$.
The fact that interacting dark energy alleviates the $H_0$ tension can also be
noticed in the results depicted by the confidence level contours in the ($\xi$, $H_0$) planes in Fig.~\ref{fig1}.
 The reason for that is the following: within the interacting dark matter-dark energy model explored here, the dark matter
density contribution is required to be smaller, as for negative $\xi$
the dark matter density will get an extra contribution
proportional to the dark energy one. Since CMB accurately constrains $\Omega_c h^2$, a larger value of $H_0$ will be
required in these scenarios, which in turn, will provide a better
agreement with the direct measurement of $H_0$ from the R16
prior. More quantitatively, notice, from the first column of Tab.~\ref{tab:pol}, that we obtain
$H_0=68.9\,\pm 1.2$ km/s/Mpc for ``Planck TTTEEE + lowTEB'' while,
without interaction and for the same combination of datasets, 
we have $H_0=67.30\,\pm0.64$ km/s/Mpc, as can be noticed from the first
column of Tab.~\ref{tab:pol_noint}. This effect reduces at
$2\sigma$ the tension of the Planck CMB anisotropy
data in a $\Lambda$CDM framework with Riess et
al. 2016~\cite{R16} (i.e. with the value $H_0=73.24\pm1.74$ km/s/Mpc). 
It should not therefore come as a surprise that when the R16 prior on the Hubble constant is added in the analyses, 
a preference for a non-zero coupling appears with a significance larger than $2$ standard deviations (see also Fig.~\ref{fig1}).  
Indeed, the "Planck TT+lowTEB+R16" dataset (see Tab.~\ref{tab:temp}) gives $\xi=-0.25\,^{+0.05}_{-0.10}$ at
$68\%$ CL ($\xi=-0.25_{-0.13}^{+0.17}$ at $95\%$ CL), while the
"Planck TTTEEE+lowTEB+R16" dataset (see Tab.~\ref{tab:pol}) gives $\xi=-0.259_{-0.098}^{+0.043}$ at
$68\%$ CL ($\xi=-0.26_{-0.12}^{+0.16}$ at $95\%$ CL).

\begin{figure*}[t]
\begin{center}
\includegraphics[width=.4\textwidth]{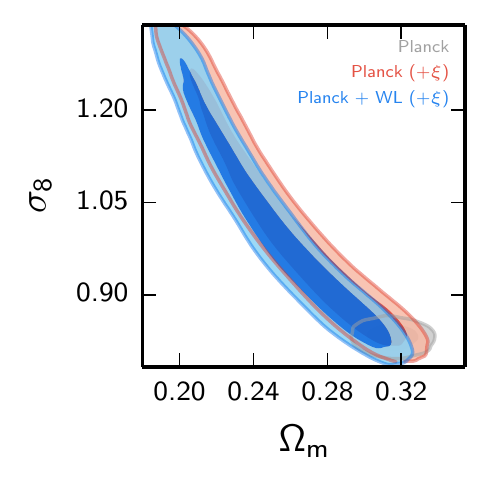}\includegraphics[width=.4\textwidth]{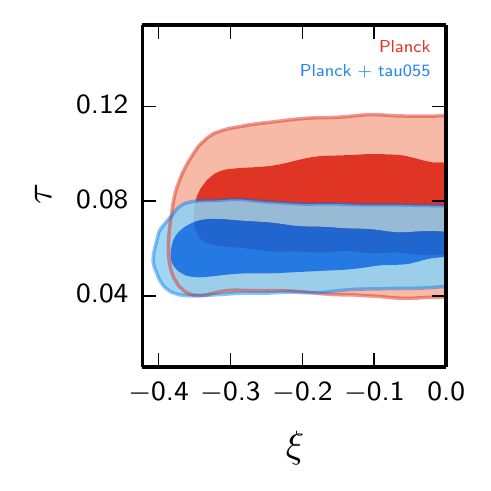}
\caption{Left panel: $68\%$ and $95\%$ CL in the two-dimensional ($\Omega_m$, $\sigma_8$) plane from the ``Planck TTTEEE + lowTEB''
dataset for a pure $\Lambda$CDM scenario, a varying $\xi$
interacting model, and also adding to the former the WL
dataset. Notice that the coupling allows for larger values
for $\sigma_8$ and smaller values for $\Omega_m$, relaxing the Planck bounds on
the $S_8$ parameter and mildly alleviating the tension with the $S_8$ values measured by cosmic shear surveys
as CFHTLenS and KiDS-450. Right panel: $68\%$ and $95\%$ CL in the two-dimensional 
($\xi$, $\tau$) plane from the  ``Planck TTTEEE '' dataset, and also combined with the "tau055" prior
on the reionization optical depth. Notice that the "tau055" prior
affects only marginally the constraints on $\xi$,
resulting in a $\sim 1 \sigma$ indication for $\xi<0$ (after marginalization over $\tau$).}
\label{fig2}
\end{center}
\end{figure*}

While the Planck+R16 combined datasets clearly show evidence for a coupling, also other
datasets seem to suggest this possibility.
When the WL dataset is included a mild indication (slightly above one standard deviation) is
indeed present. The ``Planck TT+lowTEB+WL'' dataset (see Tab.~\ref{tab:temp}) gives $\xi=-0.17\,^{+0.12}_{-0.10}$ at
$68\%$ CL, while the ``Planck TTTEEE+lowTEB+WL'' dataset 
(see Tab.~\ref{tab:pol}) gives $\xi=-0.18\,^{+0.13}_{-0.10}$ at $68\%$ CL.
It is worthwhile to note that when the interacting dark energy is
introduced, a very large shift towards lower values of the cold dark
matter density $\Omega_ch^2$ appears and the error bars are relaxed by 
a factor $10$, as can be seen by comparing the results of
Tabs.~\ref{tab:temp} and \ref{tab:pol} with those shown in
Tab.~\ref{tab:pol_noint}. Moreover, we find a shift of the
clustering parameter $\sigma_8$ towards an higher value, compensated
by a lowering of the matter density $\Omega_m$, both with relaxed
error bars. In this way, we are not increasing the tension on 
the $S_8$ parameter between Planck CMB data and the weak lensing measurements from the
CFHTLenS survey~\cite{Heymans:2012gg,Erben:2012zw} and KiDS-450~\cite{Hildebrandt:2016iqg}.
This can be also clearly noticed from the left panel of Fig.~\ref{fig2}, where we plot the two-dimensional constraints on the
$\Omega_m$-$\sigma_8$ plane from the ``Planck TTTEEE + lowTEB''
dataset in the cases of $\xi=0$, non-zero coupling, and combined with the WL measurements.

Also when the "tau055" prior is included a small preference for a
non-zero dark matter-dark energy coupling seems to emerge.
The reported constraints of $\xi=-0.21\,^{+0.09}_{-0.16}$ at $68 \%$ CL 
for the ``Planck TT+tau055'' and $\xi = -0.21\,^{+0.08}_{-0.15}$ at $68 \%$ CL for the
``Planck TTTEEE+tau055'' datasets respectively could naively suggest a statistically significant
detection (at more than $2$ standard deviations) for a dark energy-dark matter interaction.
It is however important to point out that the posterior distribution for $\xi$ is in
this case highly non gaussian. At $95 \%$ CL we have found that
the tau055 prior gives no evidence for an interaction, providing a lower
limit only. We obtain $\xi >-0.37$ for the ``Planck TT+tau055''
dataset, and $\xi >-0.38$ for the ``Planck TTTEEE+tau055''
dataset, respectively, both at $95 \%$ CL.
Therefore, while the tau055 prior suggests a value of $\xi<0$,   
this indication is not statistically significant (only slightly above
$1 \sigma$); notice this fact from the right panel of Fig.~\ref{fig2}, where we plot
the two-dimensional posteriors in the $\xi$ vs $\tau$ plane.
This preference is driven by the smaller value required in interacting dark energy models for the present dark
matter mass-energy density, which would itself lead to a lower value of $\tau$.

\begin{figure*}[t]
\begin{center}
\includegraphics[width=.49\textwidth]{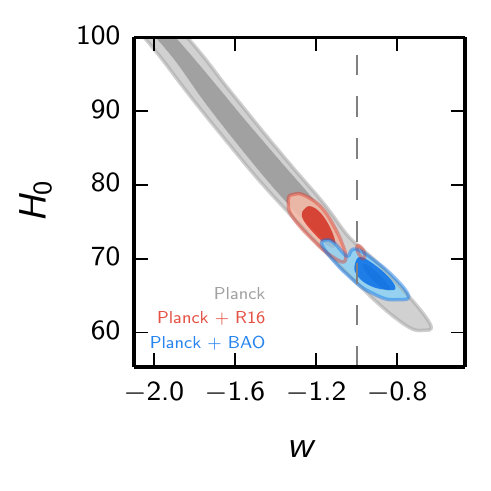}\includegraphics[width=.49\textwidth]{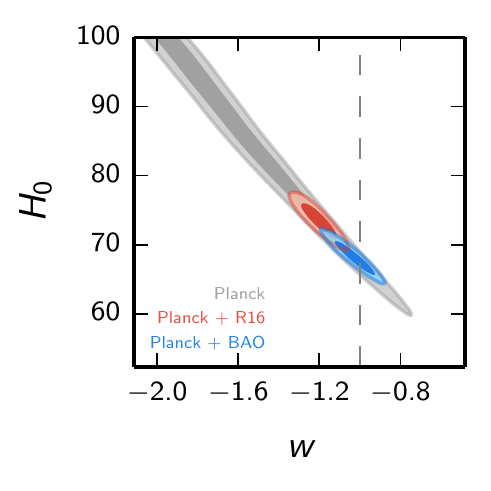}
\end{center}
\caption{Left panel: $68\%$ and $95\%$ CL in the two-dimensional 
($w$, $H_0$) plane from the combination of ``Planck TTTEEE +
lowTEB'' measurements (grey contours),  ``Planck TTTEEE +
lowTEB+ R16'' (red contours) and ``Planck TTTEEE +
lowTEB+ BAO'' (blue contours), for an interacting dark matter-dark energy
scenario. Right panel: As in the left panel but with $\xi=0$.}
\label{fig3}
\end{figure*}

In a second step, we consider the dark energy equation of state $w$ free
to vary. Indeed, if dark energy is an interacting fluid, it is
expected that its equation of state differs from the canonical value
in the $\Lambda$CDM scenario, $w=-1$. 
As we can see from the values reported in Tabs.~\ref{tab:temp_w} and ~\ref{tab:pol_w} 
and from Fig.~\ref{fig3}, where we plot the two-dimensional posteriors in the $H_0$ vs $w$ plane,
the inclusion of a negative coupling $\xi$ makes models with larger
values of $w$ in better agreement with the CMB data. Indeed, from the ``Planck TTTEEE +lowTEB'' dataset, 
we obtain the upper limit $w<-1.17$ at $68 \%$ CL when a negative coupling is
considered (see Tab.~\ref{tab:pol_w}),  
to be compared with the limit $w <-1.35$ at $68 \%$ CL obtained when we fix $\xi=0$ (see Tab.~\ref{tab:pol_noint}).
This fact results even more evident by the comparison of the
two-dimensional posteriors in the  $H_0$ vs $w$ plane depicted in Fig.~\ref{fig3} from
the ``Planck TTTEEE +lowTEB'' dataset in the case of a negative coupling
(left panel) to those with $\xi=0$ (right panel). The CMB-only
contours clearly extend to larger values of $w$ in the presence of a
negative coupling. Furthermore, the posterior in this case
appears as bimodal: there is a significant portion of models with $w>-1$ and $\xi<0$ compatible with the data.
The reason for this is simple: models with negative coupling mimic an effective equation of state with
$w<-1$. Increasing $w$ has therefore a similar effect of decreasing $\xi$ and this enhances the compatibility
of models with $w>-1$ with the data. Also from Fig.~\ref{fig3} (and as
it is well known), one can clearly noticed that when a variation in $w$ is considered, 
the Planck constraints on the Hubble constant practically vanish. In this case we can safely
include the R16 prior on the Hubble constant, as the tension between
Planck and R16 disappears, finding for this particular combination a detection of $w<-1$ at more than
$2\sigma$, obtaining $w=-1.174_{-0.072}^{+0.057}$ from "Planck TT + lowTEB+R16" (see Tab.~\ref{tab:temp_w}) and
$w=-1.184\,\pm0.064$ from "Planck TTTEEE + lowTEB+R16" (see Tab.~\ref{tab:pol_w}), both at $68 \%$ CL

When a variation on $w$ is included, the previous hint for
$\xi<0$ from the Planck+R16 dataset is still present but relaxed.
In fact, there is still a mild indication at about $1\sigma$ for
$\xi <0$ from the ``Planck TT +R16'' dataset that persists when the Planck polarization data
is included: we obtain $\xi=-0.17\,^{+0.12}_{-0.09}$ from "Planck TT+lowTEB+R16" (see Tab.~\ref{tab:temp_w})  
and  $\xi=-0.16\,^{+0.14}_{-0.06}$ from "Planck TTTEEE+lowTEB+R16" (see Tab.~\ref{tab:pol_w}), both  at $68 \%$
CL.  As we can see from Fig.~\ref{fig3} the contour plots in the case of "Planck TTTEEE+lowTEB+R16" (left panel) 
fully confirm the preference for $w<-1$ but also appear as slightly bimodal. The reason is that, for stability reasons, 
we have not considered models with positive coupling $\xi$ that would have been degenerate with models with $w<-1$.
Once again, introducing models with negative coupling increases the
compatibility with the data of models in which $w>-1$, however
in case of the $R16$ prior this is not enough to prevent an indication for $w<-1$ at more than $95 \%$ CL.

When either BAO or JLA measurements are added in the analyses the
indication for $w<-1$ disappears and a cosmological constant is now consistent
with the data within two standard deviations.
It is interesting to note, from Tab.~\ref{tab:temp_w}, that in the
case of ``Planck TT+lowTEB+BAO'' and ``Planck TTTEEE+lowTEB+BAO'' datasets one gets 
the constraint $\xi=-0.26\,^{+0.09}_{-0.18}$ at $68 \%$ CL, apparently suggesting
the evidence for a dark matter-dark energy coupling at more than $2\sigma$. 
This is probably due to the small tension present between the Planck and BAO data. 
We have found however that also in this case the posterior for $\xi$ is highly non-gaussian 
 and that the indication for a coupling from this dataset is only slightly larger than
one standard deviation. Indeed, if we consider the two standard deviation constraint
we obtain only a lower limit ($\xi>-0.426$ at $95 \%$ CL).
Since a negative coupling is degenerate with models in which $w>-1$,
we find that, when the BAO or JLA datasets are included, the constraints on
$w$ are shifted towards larger values with respect to the case with no coupling, hinting to $w>-1$ at more than one standard deviation.
Indeed, we get $w=-0.918^{+0.076}_{-0.062}$ from "Planck TT+lowTEB+BAO'' (see Tab.~\ref{tab:temp_w}) 
and $w=-0.934^{+0.071}_{-0.054}$ from "Planck TTTEEE+lowTEB+BAO'' (see
Tab.~\ref{tab:pol_w}), both at $68 \%$ CL 
to be compared to the value $w=-1.030^{+0.070}_{-0.058}$ from "Planck TTTEEE+lowTEB+BAO'' at $68 \%$ CL
but with no coupling (see Tab.~\ref{tab:pol_noint}).
Similarly, we get $w=-0.932\pm-0.067$ from "Planck TT+lowTEB+JLA'' (see Tab.~\ref{tab:temp_w}) 
and $w=-0.934\pm0.064$ from "Planck TTTEEE+lowTEB+JLA'' (see
Tab.~\ref{tab:pol_w}) at $68 \%$ CL, to be compared to the value
$w=-1.034\pm0.053$ obtained from "Planck TTTEEE+lowTEB+JLA'' at $68 \%$ CL
but with $\xi=0$ (see Tab.~\ref{tab:pol_noint}).

Within the $w$CDM +
$\xi$ scenario, the shift towards lower values of the cold dark
matter density $\Omega_ch^2$ is incremented, as can be noticed 
by comparing the results in Tab.~\ref{tab:pol_w} with those in
Tab.~\ref{tab:pol_noint}, and therefore the tension between the Planck values and
the weak lensing estimations of the $S_8$ parameter gets alleviated.
When the WL dataset is included one gets an indication for a negative 
coupling and $w<-1$ at slightly more than one standard deviation.

\begin{table*}
\begin{center}\footnotesize
\scalebox{0.8}{\begin{tabular}{c|ccccccc}
Parameter         &Planck TT& Planck TT& Planck TT & Planck TT & Planck TT & Planck TT & Planck TT  \\            
 & + lowTEB & + lowTEB + R16 & + lowTEB + BAO& + lowTEB + JLA & + lowTEB + WL & + lowTEB + lensing & + tau055 \\                 
\hline
\hspace{1mm}\\
$\Omega_bh^2$  &  $0.02226\,\pm0.00023$& $0.02223\pm0.00022$  & $0.02228\pm0.0022$& $0.02223\pm0.00023$ &  $0.02235\,\pm0.00023$& $0.02228\pm0.00023$  & $0.02211\pm0.00022$ \\
\hspace{1mm}\\
$\Omega_ch^2$  &  $0.100\,^{+0.016}_{-0.010}$&  $0.100\,^{+0.015}_{-0.010}$&  $0.088\,^{+0.012}_{-0.022}$&  $ 0.089\,^{+0.012}_{-0.022}$ &  $0.101\,^{+0.014}_{-0.008}$&  $0.097\,^{+0.017}_{-0.010}$&  $0.099\,^{+0.020}_{-0.010}$\\
\hspace{1mm}\\
$\tau$ &   $0.074\,\pm0.020$& $0.074\,\pm0.019$& $0.079\pm0.0019$ &   $0.076\,\pm0.020$& $0.073\,\pm0.020$&  $0.060\pm0.018$& $0.0580\pm0.0087$ \\
\hspace{1mm}\\
$n_s$ &  $0.9657\,\pm0.0062$  & $0.9648\,\pm0.0061$&$0.9672\,\pm0.0056$ &  $0.9656\,\pm0.0062$& $0.9692\,\pm0.0060$  & $0.9682\,\pm0.0061$&$0.9589\,\pm0.0059$  \\
\hspace{1mm}\\
$ln(10^{10}A_s)$ &  $3.082\,\pm0.038$& $3.082\,\pm0.036$& $3.091\,\pm0.037$ & $3.086\,\pm0.038$& $3.076\,\pm0.038$& $3.050\,\pm0.033$& $3.055\,\pm0.018$\\
\hspace{1mm}\\
$H_0 [\rm{km \, s^{-1} \, Mpc^{-1}}]$ &$82\,^{+10}_{-8}$&$74.2\,^{+2.2}_{-1.8}$ &$67.7\,^{+1.3}_{-1.6}$ &$67.9\,\pm1.5$ &$86\,^{+10}_{-5}$&$79\,\pm10$ &$78\,^{+10}_{-20}$  \\
\hspace{1mm}\\
$\sigma_8$ &$1.12\,_{-0.09}^{+0.17}$& $1.06\,_{-0.15}^{+0.09}$& $1.08\,\pm0.15$ &$1.09\,\pm0.15$& $1.12\,^{+0.14}_{-0.10}$  & $1.09\,_{-0.12}^{+0.18}$& $1.11\,^{+0.18}_{-0.10}$ \\
\hspace{1mm}\\
$\Omega_m$ & $0.190\,^{+0.022}_{-0.059}$  & $0.223\,\pm0.025$& $0.243\,^{+0.027}_{-0.049}$ &$0.242\,_{-0.048}^{+0.029}$& $0.174\,^{+0.017}_{-0.047}$  & $0.199\,_{-0.70}^{+0.28}$& $0.212\,^{+0.032}_{-0.076}$ \\
\hspace{1mm}\\
$\xi$  &$>-0.194$&     $-0.17\,^{+0.12}_{-0.09}$&$-0.26\,^{+0.09}_{-0.18}$  &            $-0.27\,^{+0.09}_{-0.18}$&$-0.13\,^{+0.12}_{-0.04}$&            $-0.17\,^{+0.16}_{-0.05}$&$>-0.241$\\
\hspace{1mm}\\
$w$ &  $-1.39\,^{+0.24}_{-0.41}$& $-1.174\,^{+0.057}_{-0.072}$ &$-0.918\,^{+0.076}_{-0.062}$&         $-0.932\,\pm0.067$ &$-1.48\,^{+0.20}_{-0.33}$ & $-1.29\,\pm0.33$ &$-1.29\,\pm0.38$ \\
\hline
\end{tabular}}
\end{center}
\caption{$68\%$~CL constraints on cosmological parameters from the 
  ``Planck TT + lowTEB'' baseline dataset and its further combinations with several datasets (see text). An interacting
  dark energy model with negative coupling $\xi\le0$ and a dark energy equation of state $w$ are assumed. Please note that in case of the "tau055" prior the "lowTEB" dataset is not considered.}
\label{tab:temp_w}
\end{table*}

\begin{table*}
\begin{center}\footnotesize
\scalebox{0.8}{\begin{tabular}{c|ccccccc}
Parameter         & Planck TTTEEE & Planck TTTEEE & Planck TTTEEE & Planck TTTEEE & Planck TTTEEE &  Planck TTTEEE & Planck TTTEEE   \\            
 & + lowTEB & + lowTEB + R16 & + lowTEB + BAO& + lowTEB + JLA & + lowTEB + WL & + lowTEB + lensing & + tau055 \\                 
\hline
\hspace{1mm}\\
$\Omega_bh^2$  &  $0.02225\,\pm0.00016$& $0.02223\pm0.00016$  & $0.02226\pm0.00015$& $0.02224\pm0.00016$ &  $0.02231\,\pm0.00016$& $0.02224\pm0.00016$  & $0.02218\pm0.00015$ \\
\hspace{1mm}\\
$\Omega_ch^2$  &  $0.100\,^{+0.016}_{-0.009}$&  $0.101\,^{+0.013}_{-0.011}$&  $0.090\,^{+0.013}_{-0.022}$&  $ 0.089\,^{+0.011}_{-0.022}$ &  $0.102\,^{+0.013}_{-0.008}$&  $0.097\,^{+0.018}_{-0.011}$&  $0.099\,^{+0.019}_{-0.010}$\\
\hspace{1mm}\\
$\tau$ &   $0.074\,\pm0.017$& $0.075\,\pm0.017$& $0.078\pm0.016$ &   $0.078\,\pm0.017$& $0.070\,\pm0.017$&  $0.054\pm0.015$& $0.0597\pm0.0089$ \\
\hspace{1mm}\\
$n_s$ &  $0.9644\,\pm0.0048$  & $0.9636\,\pm0.0045$&$0.9651\,\pm0.0044$ &  $0.9644\,\pm0.0049$& $0.9660\,\pm0.0047$  & $0.9650\,\pm0.0048$&$0.9601\,\pm0.0046$  \\
\hspace{1mm}\\
$ln(10^{10}A_s)$ &  $3.082\,\pm0.033$& $3.085\,\pm0.033$& $3.090\,\pm0.034$ & $3.091\,\pm0.033$& $3.072\,\pm0.033$& $3.040\,^{+0.027}_{-0.031}$& $3.058\,\pm0.018$\\
\hspace{1mm}\\
$H_0 [\rm{km \, s^{-1} \, Mpc^{-1}}]$ &$82\,^{+10}_{-8}$&$74.4\,^{+2.1}_{-1.7}$ &$68.0\,^{+1.3}_{-1.6}$ &$67.9\,\pm1.4$ &$86\,^{+10}_{-6}$&$79\,^{+10}_{-20}$ &$79\,\pm10$  \\
\hspace{1mm}\\
$\sigma_8$ &$1.13\,_{-0.09}^{+0.16}$& $1.05\,_{-0.14}^{+0.08}$& $1.08\,_{-0.19}^{+0.14}$ &$1.09\,\pm0.15$& $1.13\,^{+0.15}_{-0.09}$  & $1.09\,_{-0.11}^{+0.18}$& $1.11\,^{+0.17}_{-0.11}$ \\
\hspace{1mm}\\
$\Omega_m$ & $0.188\,^{+0.023}_{-0.055}$  & $0.224\,\pm0.026$& $0.244\,^{+0.027}_{-0.047}$ &$0.244\,_{-0.049}^{+0.026}$& $0.173\,^{+0.017}_{-0.045}$  & $0.201\,_{-0.069}^{+0.029}$& $0.203\,^{+0.029}_{-0.067}$ \\
\hspace{1mm}\\
$\xi$  &$-0.16\,^{+0.14}_{-0.06}$& $-0.16\,^{+0.11}_{-0.09}$&$-0.26\,^{+0.09}_{-0.18}$  &$>-0.351$&$-0.13\,^{+0.11}_{-0.05}$&            $-0.19\,^{+0.17}_{-0.06}$&$>-0.222$\\
\hspace{1mm}\\
$w$ &  $-1.41\,^{+0.24}_{-0.40}$& $-1.184\,\pm0.064$ &$-0.934\,^{+0.071}_{-0.054}$& $-0.934\,\pm0.064$ &$-1.52\,^{+0.20}_{-0.32}$ & $-1.28\,\pm0.35$ &$-1.33\,\pm0.35$ \\
\hline
\end{tabular}}
\end{center}
\caption{$68\%$~CL constraints on cosmological parameters from the 
  ``Planck TTTEEE + lowTEB'' baseline dataset and its further combinations with several datasets (see text). An interacting
  dark energy model with negative coupling $\xi\le0$ and a dark energy equation of state $w$ are assumed. Please note that in case of the "tau055" prior the "lowTEB" dataset is not considered.}
\label{tab:pol_w}
\end{table*}

\begin{table*}
\begin{center}\footnotesize
\scalebox{0.7}{\begin{tabular}{c|cccccccc}
Parameter         & Planck TTTEEE & Planck TTTEEE & Planck TTTEEE & Planck TTTEEE & Planck TTTEEE &  Planck TTTEEE & Planck TTTEEE & Planck TTTEEE   \\            
 & + lowTEB &  + lowTEB + BAO& + lowTEB + JLA & + lowTEB + lensing & + lowTEB &  + lowTEB + BAO& + lowTEB + JLA & + lowTEB + lensing  \\                 
\hline
\hspace{1mm}\\
$\Omega_bh^2$  &  $0.02226\,\pm0.00015$& $0.02229\pm0.00014$  & $0.02227\pm0.00015$& $0.02226\pm0.00016$ &  $0.02228\,\pm0.00016$& $0.02227\pm0.00015$  & $0.02225\pm0.00016$& $0.02228\pm0.00016$ \\
\hspace{1mm}\\
$\Omega_ch^2$  &  $0.1198\,\pm0.0014$&  $0.1193\,\pm 0.0011$&  $ 0.1196\,\pm0.0014$&  $ 0.1193\,\pm 0.0014$ &  $0.1196\,\pm 0.0015$&  $0.1196\,\pm 0.0014$&  $0.1198\,\pm 0.0015$&  $0.1191\,\pm 0.0014$\\
\hspace{1mm}\\
$\tau$ &   $0.079\,\pm0.017$& $0.082\,\pm0.016$& $0.080\pm0.017$ &   $0.062\,\pm0.014$& $0.076\,\pm0.017$&  $0.080\pm0.017$& $0.079\pm0.017$& $0.056\pm0.015$ \\
\hspace{1mm}\\
$n_s$ &  $0.9646\,\pm0.0047$  & $0.9661\,\pm0.0041$&$0.9650\,\pm0.0047$ &  $0.9652\,\pm0.0048$& $0.9649\,\pm0.0047$  & $0.9651\,\pm0.0046$&$0.9645\,\pm0.0047$&$0.9657\,\pm0.0047$  \\
\hspace{1mm}\\
$ln(10^{10}A_s)$ &  $3.094\,\pm0.034$& $3.098\,\pm 0.032$& $3.094\,\pm0.033$ & $3.058\,\pm0.025$& $3.086\,\pm0.034$& $3.095\,\pm0.034$& $3.093\,\pm0.033$& $3.044\,\pm0.028$\\
\hspace{1mm}\\
$H_0 [\rm{km \, s^{-1} \, Mpc^{-1}}]$ &$67.30\,\pm0.64$&$67.53\,\pm 0.48$ &$67.36\,\pm0.64$ &$67.50\,\pm0.64$ &$>80.9$&$68.2\,^{+1.4}_{-1.7}$ &$68.3\,\pm1.6$&$>75.3$  \\
\hspace{1mm}\\
$\sigma_8$ &$0.831\,_{-0.013}^{+0.015}$& $0.832\,\pm 0.013$& $0.831\,\pm0.013$ &$0.8146\,\pm0.0088$& $0.98\,^{+0.11}_{-0.06}$  & $0.839\,\pm 0.022$& $0.841\,\pm0.020$& $0.924\,^{+0.12}_{-0.07}$ \\
\hspace{1mm}\\
$\Omega_m$ & $0.3152\,\pm 0.0089$  & $0.3119\,\pm 0.0065$& $0.3142\,\pm 0.0089$ &$0.3122\,_{-0.0095}^{+0.0084}$& $0.205\,^{+0.023}_{-0.066}$  & $0.307\,\pm 0.013$& $0.307\,^{+0.014}_{-0.017}$& $0.227\,^{+0.036}_{-0.089}$ \\
\hspace{1mm}\\
$w$ &  $[-1]$& $[-1]$ &$[-1]$& $[-1]$ &$-1.54\,^{+0.19}_{-0.38}$ & $-1.030\,^{+0.070}_{-0.058}$ &$-1.034\,\pm0.053$&$-1.42\,^{+0.27}_{-0.45}$ \\
\hline
\end{tabular}}
\end{center}
\caption{$68\%$~CL constraints on cosmological parameters from the 
  ``Planck TTTEEE + lowTEB'' baseline dataset and its combination with external datasets assuming $\Lambda$CDM and
  $w$CDM models in the absence of an interaction, i.e. $\xi=0$.}
\label{tab:pol_noint}
\end{table*}

\section{Conclusions}

We have explored the well-known Hubble constant $H_0$ tension between the current estimates from
late-time universe data of Riess et al. 2016, which indicates a value
of $H_0=73.24\pm1.74$ km/s/Mpc, and the Planck Cosmic Microwave
Background (CMB) measurement of $H_0=66.93 \pm 0.62$ km/s/Mpc, (both at $68
\%$ CL), in the context of interacting dark matter-dark energy
scenarios. Such a coupling could affect the value of the present matter 
energy density $\Omega_m$. Therefore, if within an interacting
model $\Omega_m$ is smaller, a larger value of $H_0$ would be
required in order to satisfy the peaks structure of CMB observations, which accurately
determine the value of $\Omega_mh^2$. We find that for one of the most interesting and viable
coupled dark matter-dark energy scenarios in the literature, in which the exchanged energy rate is negative (i.e. the energy flows
from the dark matter system to the dark energy one) and proportional
to the dark energy density, the existing $3\sigma$ $H_0$ tension is
alleviated. In addition, when combining CMB measurements with the
Hubble constant prior from Riess et al. 2016, a preference for a
non-zero coupling appears with a significance larger than $2\sigma$.
However, it is certainly not unexpected that in interacting scenarios
the dark energy equation of state differs from its canonical value
within the $\Lambda$CDM picture, i.e., is different from $w=-1$. We
have therefore considered as well such a possibility, finding that,
when the dark energy equation of state $w$ is also a free parameter,
the Hubble constant tension gets strongly alleviated, obtaining $\sim
3\sigma$ evidence for a phantom-like ($w<-1$) dark energy fluid when
combining CMB and Riess et al. 2016 measurements. 
However, when other datasets, as BAO or Supernovae Ia luminosity distances from JLA, 
are also included in our numerical analyses, a good consistency with a pure $\Lambda$CDM
cosmological scenario is found with models with negative coupling
and $w>-1$ suggested at slightly more than one standard deviation.

\begin{acknowledgments}
We would like to thank Laura Lopez-Honorez for stimulating discussions
and help with the numerical codes. 
The work done by AM is also supported by TASP/INFN. 
O.M. is supported by PROMETEO II/2014/050, by the Spanish
Grant FPA2014--57816-P of the MINECO, by the MINECO Grant
SEV-2014-0398 and by the European Union's Horizon 2020
research and innovation programme under the Marie Sklodowska-Curie grant
agreements 690575 and 674896. 
This work has been done within the Labex ILP (reference ANR-10-LABX-63) part of the Idex SUPER, and received financial state aid managed by the Agence Nationale de la Recherche, as part of the programme Investissements d'avenir under the reference ANR-11-IDEX-0004-02. 
\end{acknowledgments}


\begin{thebibliography}{69}%
\bibitem{planckparams2015}
P.~A.~R.~Ade {\it et al.} [Planck Collaboration],
  Astron.\ Astrophys.\  {\bf 594}, A13 (2016)
  doi:10.1051/0004-6361/201525830
  [arXiv:1502.01589 [astro-ph.CO]].
  
\bibitem{newtau}
 N.~Aghanim {\it et al.} [Planck Collaboration],
  Astron.\ Astrophys.\  {\bf 596}, A107 (2016)
  doi:10.1051/0004-6361/201628890
  [arXiv:1605.02985 [astro-ph.CO]].

\bibitem{R16}
  A.~G.~Riess {\it et al.},
  arXiv:1604.01424 [astro-ph.CO].

\bibitem{Heymans:2012gg} 
  C.~Heymans {\it et al.},
  Mon.\ Not.\ Roy.\ Astron.\ Soc.\  {\bf 427}, 146 (2012)
  doi:10.1111/j.1365-2966.2012.21952.x
  [arXiv:1210.0032 [astro-ph.CO]].
\bibitem{Erben:2012zw} 
  T.~Erben {\it et al.},
  Mon.\ Not.\ Roy.\ Astron.\ Soc.\  {\bf 433}, 2545 (2013)
  doi:10.1093/mnras/stt928
  [arXiv:1210.8156 [astro-ph.CO]].
\bibitem{Hildebrandt:2016iqg} 
  H.~Hildebrandt {\it et al.},
  arXiv:1606.05338 [astro-ph.CO].
\bibitem{Joudaki:2016kym} 
  S.~Joudaki {\it et al.},
  arXiv:1610.04606 [astro-ph.CO].

\bibitem{dms0}
  E.~Di Valentino, A.~Melchiorri and J.~Silk,
  Phys.\ Rev.\ D {\bf 92} (2015) no.12,  121302
  doi:10.1103/PhysRevD.92.121302
  [arXiv:1507.06646 [astro-ph.CO]].

\bibitem{dms}
 E.~Di Valentino, A.~Melchiorri and J.~Silk,
  Phys.\ Lett.\ B {\bf 761} (2016) 242
  doi:10.1016/j.physletb.2016.08.043
  [arXiv:1606.00634 [astro-ph.CO]].
  \bibitem{bernal}
  J.~L.~Bernal, L.~Verde and A.~G.~Riess,
  JCAP {\bf 1610} (2016) no.10,  019
  doi:10.1088/1475-7516/2016/10/019
  [arXiv:1607.05617 [astro-ph.CO]].
\bibitem{dmnu1}
  R.~J.~Wilkinson, C.~Boehm and J.~Lesgourgues,
  JCAP {\bf 1405} (2014) 011
  doi:10.1088/1475-7516/2014/05/011
  [arXiv:1401.7597 [astro-ph.CO]].
\bibitem{dmnu2}
\ J.~Lesgourgues, G.~Marques-Tavares and M.~Schmaltz,
  JCAP {\bf 1602} (2016) no.02,  037
  doi:10.1088/1475-7516/2016/02/037
  [arXiv:1507.04351 [astro-ph.CO]].
\bibitem{decay}
  Z.~Berezhiani, A.~D.~Dolgov and I.~I.~Tkachev,
  Phys.\ Rev.\ D {\bf 92} (2015) no.6,  061303
  doi:10.1103/PhysRevD.92.061303
  [arXiv:1505.03644 [astro-ph.CO]].
\bibitem{Gariazzo:2017pzb} 
  S.~Gariazzo, M.~Escudero, R.~Diamanti and O.~Mena,
  arXiv:1704.02991 [astro-ph.CO].
  
\bibitem{Grandis:2016fwl}
  S.~Grandis, D.~Rapetti, A.~Saro, J.~J.~Mohr and J.~P.~Dietrich,
  Mon.\ Not.\ Roy.\ Astron.\ Soc.\  {\bf 463} (2016) no.2,  1416
  doi:10.1093/mnras/stw2028
  [arXiv:1604.06463 [astro-ph.CO]].

\bibitem{Zhao:2017urm}
  M.~M.~Zhao, D.~Z.~He, J.~F.~Zhang and X.~Zhang,
arXiv:1703.08456 [astro-ph.CO].

\bibitem{Yang:2017amu}
  W.~Yang, R.~C.~Nunes, S.~Pan and D.~F.~Mota,
arXiv:1703.02556 [astro-ph.CO].

\bibitem{Prilepina:2016rlq}
  V.~Prilepina and Y.~Tsai,
arXiv:1611.05879 [hep-ph].

\bibitem{DiValentino:2016ziq} 
  E.~Di Valentino and L.~Mersini-Houghton,
  JCAP {\bf 1703}, no. 03, 020 (2017)
  doi:10.1088/1475-7516/2017/03/020
  [arXiv:1612.08334 [astro-ph.CO]].

\bibitem{Santos:2016sog}
  B.~Santos, A.~A.~Coley, N.~C.~Devi and J.~S.~Alcaniz,
JCAP {\bf 1702} (2017) no.02,  047
doi:10.1088/1475-7516/2017/02/047
[arXiv:1611.01885 [astro-ph.CO]].

\bibitem{Kumar:2016zpg}
  S.~Kumar and R.~C.~Nunes,
Phys.\ Rev.\ D {\bf 94} (2016) no.12,  123511
doi:10.1103/PhysRevD.94.123511
[arXiv:1608.02454 [astro-ph.CO]].


\bibitem{Karwal:2016vyq} 
  T.~Karwal and M.~Kamionkowski,
Phys.\ Rev.\ D {\bf 94}, no. 10, 103523 (2016)
doi:10.1103/PhysRevD.94.103523
[arXiv:1608.01309 [astro-ph.CO]].

\bibitem{Ko:2016uft}
  P.~Ko and Y.~Tang,
Phys.\ Lett.\ B {\bf 762} (2016) 462
doi:10.1016/j.physletb.2016.10.001
[arXiv:1608.01083 [hep-ph]].

\bibitem{Archidiacono:2016kkh} 
  M.~Archidiacono, S.~Gariazzo, C.~Giunti, S.~Hannestad, R.~Hansen, M.~Laveder and T.~Tram,
JCAP {\bf 1608}, no. 08, 067 (2016)
doi:10.1088/1475-7516/2016/08/067
[arXiv:1606.07673 [astro-ph.CO]].

\bibitem{Qing-Guo:2016ykt}
  Q.~G.~Huang and K.~Wang,
Eur.\ Phys.\ J.\ C {\bf 76} (2016) no.9,  506
doi:10.1140/epjc/s10052-016-4352-x
[arXiv:1606.05965 [astro-ph.CO]].

\bibitem{Chacko:2016kgg}
  Z.~Chacko, Y.~Cui, S.~Hong, T.~Okui and Y.~Tsai,
  JHEP {\bf 1612} (2016) 108
  doi:10.1007/JHEP12(2016)108
  [arXiv:1609.03569 [astro-ph.CO]].
  
\bibitem{Zhang:2017idq}
  Y.~Zhang, H.~Zhang, D.~Wang, Y.~Qi, Y.~Wang and G.~B.~Zhao,
arXiv:1703.08293 [astro-ph.CO].


\bibitem{zhao}
G.~B.~Zhao {\it et al.},
  arXiv:1701.08165 [astro-ph.CO].
  
 \bibitem{Sola:2017jbl}
  J.~Sola, J.~d.~C.~Perez and A.~Gomez-Valent,
  arXiv:1703.08218 [astro-ph.CO].
  
\bibitem{brust}
  C.~Brust, Y.~Cui and K.~Sigurdson,
  arXiv:1703.10732 [astro-ph.CO].

\bibitem{Lancaster:2017ksf}
  L.~Lancaster, F.~Y.~Cyr-Racine, L.~Knox and Z.~Pan,
  arXiv:1704.06657 [astro-ph.CO].
  
\bibitem{dms02}
  E.~Di Valentino, A.~Melchiorri, E.~V.~Linder and J.~Silk,
  arXiv:1704.00762 [astro-ph.CO].


\bibitem{Salvatelli} V.~Salvatelli, N.~Said, M.~Bruni, A.~Melchiorri, and D.~Wands,
  Phys.\ Rev.\ Lett.\  {\bf 113}, 181301 (2014). [arXiv:1406.7297]

\bibitem{Sola1}  J. Sola, A. G. Valent and J. C. Perez, Astrophys. J. Lett. {\bf811}, L14 (2015). [arXiv:1506.05793] 
\bibitem{Sola2}  J. Sola, A. G. Valent and J. C. Perez, arXiv:1602.02103.

\bibitem{Saulo}  C. Pigozzo et al., JCAP \textbf{1605}, 022 (2016). [arXiv:1510.01794]

\bibitem{Richarte} M.~G.~Richarte, and L.~Xu, arXiv:1506.02518.

\bibitem{Valiviita} J.~V\"{a}liviita, and E.~Palmgren,  JCAP {\bf 1507}, 015 (2015). [arXiv:1504.02464]

\bibitem{Elisa} E. G. M. Ferreira et al., arXiv:1412.2777.


\bibitem{Murgia:2016ccp} 
  R.~Murgia, S.~Gariazzo and N.~Fornengo,
  JCAP {\bf 1604}, no. 04, 014 (2016)
  doi:10.1088/1475-7516/2016/04/014
  [arXiv:1602.01765 [astro-ph.CO]].

\bibitem{ideolga}
M.~B.~Gavela, D.~Hernandez, L.~Lopez Honorez, O.~Mena and S.~Rigolin,
  JCAP {\bf 0907}, 034 (2009)
  Erratum: [JCAP {\bf 1005}, E01 (2010)]
  doi:10.1088/1475-7516/2010/05/E01, 10.1088/1475-7516/2009/07/034
  [arXiv:0901.1611 [astro-ph.CO]];
 M.~B.~Gavela, L.~Lopez Honorez, O.~Mena and S.~Rigolin,
  JCAP {\bf 1011}, 044 (2010)
  doi:10.1088/1475-7516/2010/11/044
  [arXiv:1005.0295 [astro-ph.CO]].
\bibitem{ide2}
  V.~Salvatelli, A.~Marchini, L.~Lopez-Honorez and O.~Mena,
  Phys.\ Rev.\ D {\bf 88} (2013) no.2,  023531
  doi:10.1103/PhysRevD.88.023531
  [arXiv:1304.7119 [astro-ph.CO]].
\bibitem{riess11}
A.~G.~Riess {\it et al.},
  Astrophys.\ J.\  {\bf 730}, 119 (2011)
  Erratum: [Astrophys.\ J.\  {\bf 732}, 129 (2011)]
  doi:10.1088/0004-637X/732/2/129, 10.1088/0004-637X/730/2/119
  [arXiv:1103.2976 [astro-ph.CO]].
\bibitem{Doran:2003xq} 
  M.~Doran, C.~M.~Muller, G.~Schafer and C.~Wetterich,
  Phys.\ Rev.\ D {\bf 68}, 063505 (2003)
  doi:10.1103/PhysRevD.68.063505
  [astro-ph/0304212].
\bibitem{Majerotto:2009np} 
  E.~Majerotto, J.~Valiviita and R.~Maartens,
  Mon.\ Not.\ Roy.\ Astron.\ Soc.\  {\bf 402}, 2344 (2010)
  doi:10.1111/j.1365-2966.2009.16140.x
  [arXiv:0907.4981 [astro-ph.CO]].
\bibitem{Ballesteros:2010ks} 
  G.~Ballesteros and J.~Lesgourgues,
  JCAP {\bf 1010}, 014 (2010)
  doi:10.1088/1475-7516/2010/10/014
  [arXiv:1004.5509 [astro-ph.CO]].
\bibitem{Aghanim:2015xee} 
  N.~Aghanim {\it et al.} [Planck Collaboration],
  Astron.\ Astrophys.\  {\bf 594}, A11 (2016)
  doi:10.1051/0004-6361/201526926
  [arXiv:1507.02704 [astro-ph.CO]].
\bibitem{Ade:2015zua} 
  P.~A.~R.~Ade {\it et al.} [Planck Collaboration],
  Astron.\ Astrophys.\  {\bf 594}, A15 (2016)
  doi:10.1051/0004-6361/201525941
  [arXiv:1502.01591 [astro-ph.CO]].
\bibitem{beutler2011}
  F.~Beutler
{\it et al.},
  Mon.\ Not.\ Roy.\ Astron.\ Soc.\  {\bf 416} (2011) 3017
  [arXiv:1106.3366 [astro-ph.CO]].
\bibitem{ross2014}
  A.~J.~Ross
{\it et al.},
  Mon.\ Not.\ Roy.\ Astron.\ Soc.\  {\bf 449} (2015) 835
  [arXiv:1409.3242 [astro-ph.CO]].
\bibitem{anderson2014}
  L.~Anderson {\it et al.}  [BOSS Collaboration],
  Mon.\ Not.\ Roy.\ Astron.\ Soc.\  {\bf 441} (2014) 1,  24
  [arXiv:1312.4877 [astro-ph.CO]].
\bibitem{JLA}
  M.~Betoule {\it et al.}  [SDSS Collaboration],
  Astron.\ Astrophys.\  {\bf 568} (2014) A22
  [arXiv:1401.4064 [astro-ph.CO]].

\bibitem{Lewis:2002ah} 
  A.~Lewis and S.~Bridle,
  Phys.\ Rev.\ D {\bf 66}, 103511 (2002)
  [astro-ph/0205436].
\bibitem{Lewis:2013hha} 
  A.~Lewis,
  Phys.\ Rev.\ D {\bf 87}, no. 10, 103529 (2013)
  doi:10.1103/PhysRevD.87.103529
  [arXiv:1304.4473 [astro-ph.CO]].

\end{thebibliography}
\end{document}